\documentclass[submission,copyright]{eptcs}

\usepackage[english]{babel}
\usepackage{todonotes}
\usepackage{amsmath}
\usepackage{cleveref}

\newcommand{\eg}{e.\,g.,\ }

\usepackage[acronym]{glossaries} 
\newacronym{api}{API}{Application Programming Interface}
\newacronym{ast}{AST}{Abstract Syntax Tree}
\newacronym{atp}{ATP}{Automated Theorem Proving}
\newacronym{cli}{CLI}{Command-Line Interface}
\newacronym{ct}{CT}{Combinatorial Testing}
\newacronym{czt}{CZT}{Community Z Tools}
\newacronym{dap}{DAP}{Debug Adapter Protocol}
\newacronym{dsl}{DSL}{Domain-Specific Language}
\newacronym{glsp}{GLSP}{Graphical Language Server Platform}
\newacronym{gui}{GUI}{Graphical User Interface}
\newacronym{ide}{IDE}{Integrated Development Environment}
\newacronym{itp}{ITP}{Interactive Theorem Proving}
\newacronym{loc}{LoC}{Lines of Code}
\newacronym{lpf}{LPF}{Logic of Partial Functions}
\newacronym{lsp}{LSP}{Language Server Protocol}
\newacronym{lsi}{LSI}{Language Server Interface}
\newacronym{lsif}{LSIF}{Language Server Index Format}
\newacronym{po}{PO}{Proof Obligation}
\newacronym{poc}{PoC}{Proof of Concept}
\newacronym{pog}{POG}{Proof Obligation Generation}
\newacronym{pvs}{PVS}{Prototype Verification System}
\newacronym{rcp}{RCP}{Rich Client Platform}
\newacronym{rpc}{RPC}{Remote Procedure Call}
\newacronym{slsp}{SLSP}{Specification Language Server Protocol}
\newacronym{tap}{TAP}{Test Anything Protocol}
\newacronym{tp}{TP}{Theorem Proving}
\newacronym{ui}{UI}{User Interface}
\newacronym{uri}{URI}{Uniform Resource Identifier}
\newacronym{vdm}{VDM}{Vienna Development Method}
\newacronym{vscode}{VS Code}{Visual Studio Code}
\newglossaryentry{latex}{
    name=LaTeX,
    description={Is a software system for document preparation, that is often used by academics}
}
\newcommand{\citeurl}[1]{\footnote{\url{#1}}}

\usepackage{placeins}   

\title{The Specification Language Server Protocol:\\ A Proposal for Standardised LSP Extensions}

\providecommand{\event}{F-IDE2021} % Name of the event you are submitting to

\author{
    Jonas Kjær Rask \quad\qquad Frederik Palludan Madsen
    \institute{
    DIGIT, Aarhus University, Department of Electrical and Computer Engineering, \\
    Finlandsgade 22, 8200 Aarhus N, Denmark}
    \email{jkr@ece.au.dk \qquad\qquad fpm@ece.au.dk}
\and
    Nick Battle
    \institute{Independent}
    \email{nick.battle@acm.org}
\and
    Hugo Daniel Macedo \quad\qquad Peter Gorm Larsen
    \institute{
    DIGIT, Aarhus University, Department of Electrical and Computer Engineering, \\
    Finlandsgade 22, 8200 Aarhus N, Denmark}
    \email{hdm@ece.au.dk \qquad\qquad\qquad\quad pgl@ece.au.dk}
}

\def\titlerunning{The Specification Language Server Protocol: Unifying LSP Extensions} % FIX
\def\authorrunning{J.K. Rask, F.P. Madsen, N. Battle, H.D. Macedo \& P.G. Larsen} % FIX
\begin{document}
\maketitle
%\todo[inline]{Fix running title and JKR+FPM email}

\begin{abstract}
The Language Server Protocol (LSP) changed the field of Integrated Development Environments (IDEs), as it decouples core (programming) language features functionality from editor smarts, thus lowering the effort required to extend an IDE to support a language. 
The concept is a success and has been adopted by several programming languages and beyond. This is shown by the emergence of several LSP implementations for the many programming and specification languages (languages with a focus on  modelling, reasoning, or proofs).
However, for such languages LSP has been ad-hocly extended with the additional functionalities that are typically not found for programming languages and thus not supported in LSP. 
This foils the original LSP decoupling goal, because the move towards a new IDE requires yet another re-implementation of the ad-hoc LSP extension.
In this paper we contribute with a conservative extension of LSP providing a first proposal towards a standard protocol decoupling the support of specification languages from the IDE. 
We hope our research attracts the larger community and motivates the need of a joint task-force leading to a standardised LSP extension serving the particular needs of specification languages.
\end{abstract}

\section{Introduction}
\label{sec:intro}
% Hugo will have a go at this 

Modern Integrated Development Environments (IDEs) go beyond aggregating of the various tools used by developers working with a programming language. IDEs have become platform ecosystems attracting users in search for smart features (\eg code-completion, show documentation on hovering over a primitive). The ecosystems lead to market segmentation and in some cases users stand by a single ecosystem choice, which in the extreme becomes hate for the alternative IDEs, the popularly known ``editor wars''. This paves the way to a win-win strategy where programming language maintainers and tool developers jointly gain user adoption when an IDE provides support to their language and vice-versa. However, to put the strategy in practice one hits the $M \times N$ problem \cite{Rodriguez&18}: Given $M$ editors and $N$ languages, there
are $M \times N$ language support implementations to be developed. Furthermore, each is a re-implementation which opens the way to bug introduction and inconsistent outcomes of invoking the same functionality in different IDEs.

The \gls{lsp}\citeurl{https://microsoft.github.io/language-server-protocol/} changed the field of IDEs, as it 
decouples core (programming) language features functionality from editor smarts lowering the effort needed to extend a particular IDE with support to a formal language. Thus the problem becomes to produce $M + N$ implementations, which additionally may be unified.
At the cost of implementing one server for a language, one obtains language smarts (syntax highlighting, type-checking, \ldots) support in all the IDEs supporting LSP, and an IDE 
maintainer can easily support various languages by interacting with the language tools in a standard manner. 
The additional Debug Adapter Protocol (DAP)\citeurl{https://microsoft.github.io/debug-adapter-protocol/} provides the same decoupling for executable languages with debugger tool support.

The concept is a success and has been adopted by several programming languages and beyond\citeurl{https://microsoft.github.io/language-server-protocol/implementors/servers/}, which is shown by the emergence of several LSP based support for the many programming and specification languages, which have a focus on  modelling, reasoning, or proofs (thus containing additional functionalities such as the ones depicted in \Cref{fig:ProtocolFeatureCoverage}).
However, for such languages, the LSP has mainly been extended in an ad-hoc manner with the additional functionalities that are typically not found for programming languages and thus also lacking support in LSP\footnote{\eg Alloy: \url{https://github.com/DongyuZhao/vscode-alloy} and\\Isabelle: \url{https://marketplace.visualstudio.com/items?itemName=makarius.isabelle}}\cite{Masci&19}. This foils the decoupling goal as the move towards a new IDE requires yet another re-implementation of the LSP extension.

In this paper we contribute with a conservative extension of LSP, the \emph{Specification Language Server Protocol} (SLSP), providing a first proposal towards a standard protocol decoupling the support of specification languages from the IDE, encompassing:
Combinatorial Testing \cite{Ledru02,Larsen&10c}, Translation/Code-generation \cite{Mukherjee97_2,Jorgensen&17}, Proof Obligation Generation \cite{Jones85a_2,Aichernig&97_2} and Theorem Proving \cite{Nipkow89b_2}.

We report on a proof of concept for the protocol that was carried out in the context of a MSc thesis project in \cite{Rask&21}.
The end result is an implementation of the non-programming language features of the Vienna Development Method (VDM) specification languages that together with a LSP and DAP implement VDM support in Visual Studio Code (VS Code). We hope our research attracts the larger community and congregates the community towards a common effort leading to an LSP extension with effort saving benefits. Moreover, our results confirm that non-LSP based IDEs (Eclipse Rich Client Platforms based on IDE extension plugins) are more difficult to support, given the discrepancy of language specific plugins, thus confirming the value proposed by LSP, DAP, and SLSP.

The remainder of the paper is structured as follows:
In Section~\ref{sec:decoupling}, we provide an overview of the recent protocols used to decouple IDE and language features. 
Then, Section~\ref{sec:slsp} contains a description of the SLSP protocol.
The evaluation of the protocol is described in Section~\ref{sec:pilotStudy} that contains the pilot-study and its results. 
Section~\ref{sec:relatedWork} provides an overview the various ad-hoc extensions to LSP that our protocol intends to aggregate.
At last, Section~\ref{sec:conclusion} concludes the paper, and provides an outline of the future work.

\section{Decoupling Using Language-Neutral Protocols}
\label{sec:decoupling}
% Jonas
% \begin{itemize}
%     \item Existing Language-neutral Protocols: LSP + DAP
%     \item Feature coverage using LSP + DAP
%     \item What is missing for spec.lang. support
% \end{itemize}

The general purpose of IDEs is to provide support for the development of code and specifications by providing language features such as syntax-checking, hover information and code completion.
Supporting such features for a language requires significant effort, which increases for each development tool that should support the language. 
Decoupling the feature support from the development tool allows extensive reuse as the feature support can be used for multiple development tools.
A method to achieve this decoupling is to use a communication protocol in a client-server architecture separating the development tool (client) and the language feature support (server).
Thus, a development tool only has to implement the protocol to support a given language, which can be done with little effort compared to native integration \cite{Bunder19b}. 
% Using the protocol decoupling approach reduces the $M \times N$ problem of implementing support for all languages in all development tools to a $M + N$ problem.

This section introduces the existing language-neutral protocols that can be used for decoupling the analysis support from the user interface.
These are LSP and the DAP protocols.
This is followed by an overview of the features that are not currently supported by those language-neutral protocols.

\subsection{The Language Server Protocol}
\begin{figure}[htb]
    \centering
    \includegraphics[width=0.75\textwidth]{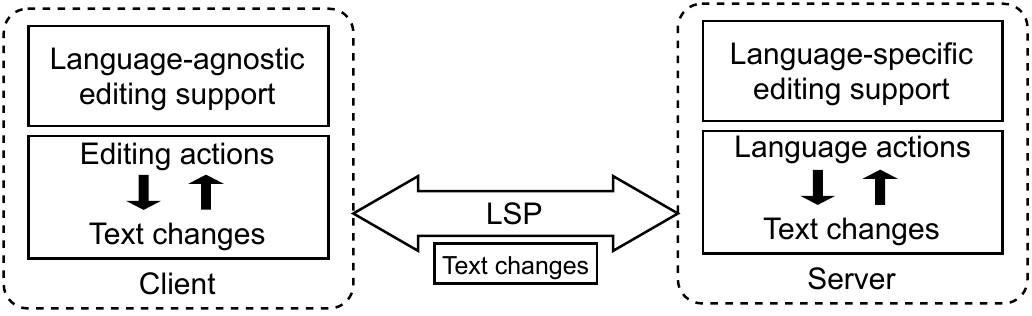}
    \caption[\acrshort{lsp} approach to language support.]
        {\acrshort{lsp} approach to language support. Borrowed from \cite{Rodriguez&18}.}
    \label{fig:LSPApproach}
\end{figure}

% \hdm{ Could we provide the idea better by adding JSON exchanges details in such a way that we can then use those in the description of how SLSP extends LSP }
% \jkr{ will have a go at this}
The LSP protocol is a standardised protocol used to decouple a language-agnostic editor (client) and a language-specific server that provides editorial language features such as syntax-checking, hover information and code completion.
This is illustrated in \Cref{fig:LSPApproach}.

The client is responsible for managing editing actions without any knowledge of the language by transmitting these to the server as text changes.
The server converts the changes into language actions, which is used to perform language-specific support and forward the information to the client.
To facilitate this the \gls{lsp} protocol uses language-neutral data types such as document references and document positions.

The base of the LSP protocol consists of a header and a content part (comparable to HTTP).
The content part of a message uses the stateless and lightweight JSON-RPC protocol\citeurl{https://www.jsonrpc.org/specification} to facilitate the three base message types: requests, responses and notifications.
JSON-RPC can be used within the same process, over sockets and many other message passing environments.
This allows the server and client to be on different physical machines.
However, most implementations run the server and client as separate process, but on the same physical machine \cite{Bunder19a}.
The request and notification messages provide the entries \texttt{method} and \texttt{params}, that specify a method (\eg `textDocument/didChange') and associated parameters.
The LSP protocol specifies a range of different messages to provide the functionality.
Furthermore, the request and response messages specify an ID, that provides easy matching of responses to requests.

\subsection{The Debug Adapter Protocol}
\begin{figure}[htb]
    \centering
        \includegraphics[width=0.8\textwidth]{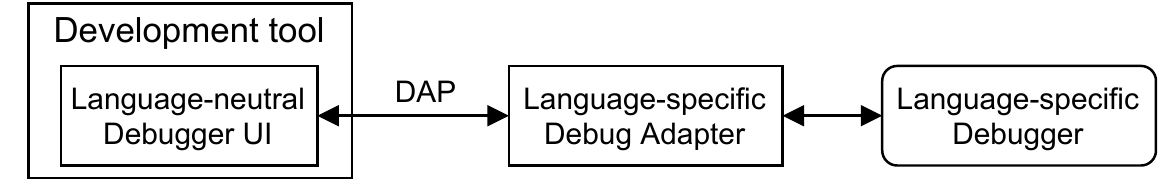}
    \caption[The decoupled architecture where the \gls{dap} protocol is used.]
        {The decoupled architecture where the \gls{dap} protocol is used. }
    \label{fig:DAPArchitecture}
\end{figure}

The DAP protocol is a standardised protocol for decoupling IDEs, editors and other development tools from the implementation of a language-specific debugger. 
The debug features supported by the protocol includes: different types of breakpoints, variable values, multi-process and thread support, navigation through data structures and more.
To be compatible with existing debugger components, the protocol relies on an intermediary debug adapter component.
It is used to wrap one or multiple debuggers, to allow communication using the DAP protocol.
The adapter is then part of a two-way communication with a language-agnostic debugger component, which is integrated into a development environment as illustrated in \Cref{fig:DAPArchitecture}. 

The DAP protocol uses a JSON-based wire-format\citeurl{https://code.visualstudio.com/blogs/2018/08/07/debug-adapter-protocol-website} inspired by the V8 Debugging Protocol\citeurl{https://github.com/dtretyakov/node-tools/wiki/Debugging-Protocol}.
This format is similar to but not compatible with the JSON-RPC used in the LSP protocol.
Otherwise, it is similar to the LSP protocol with language-neutral data types and a base protocol that has requests, responses and events (similar to notifications from the LSP protocol).

\subsection{Support for Specification Languages}
\begin{figure}[htbp]
    \centering
    \includegraphics[width=0.8\textwidth]{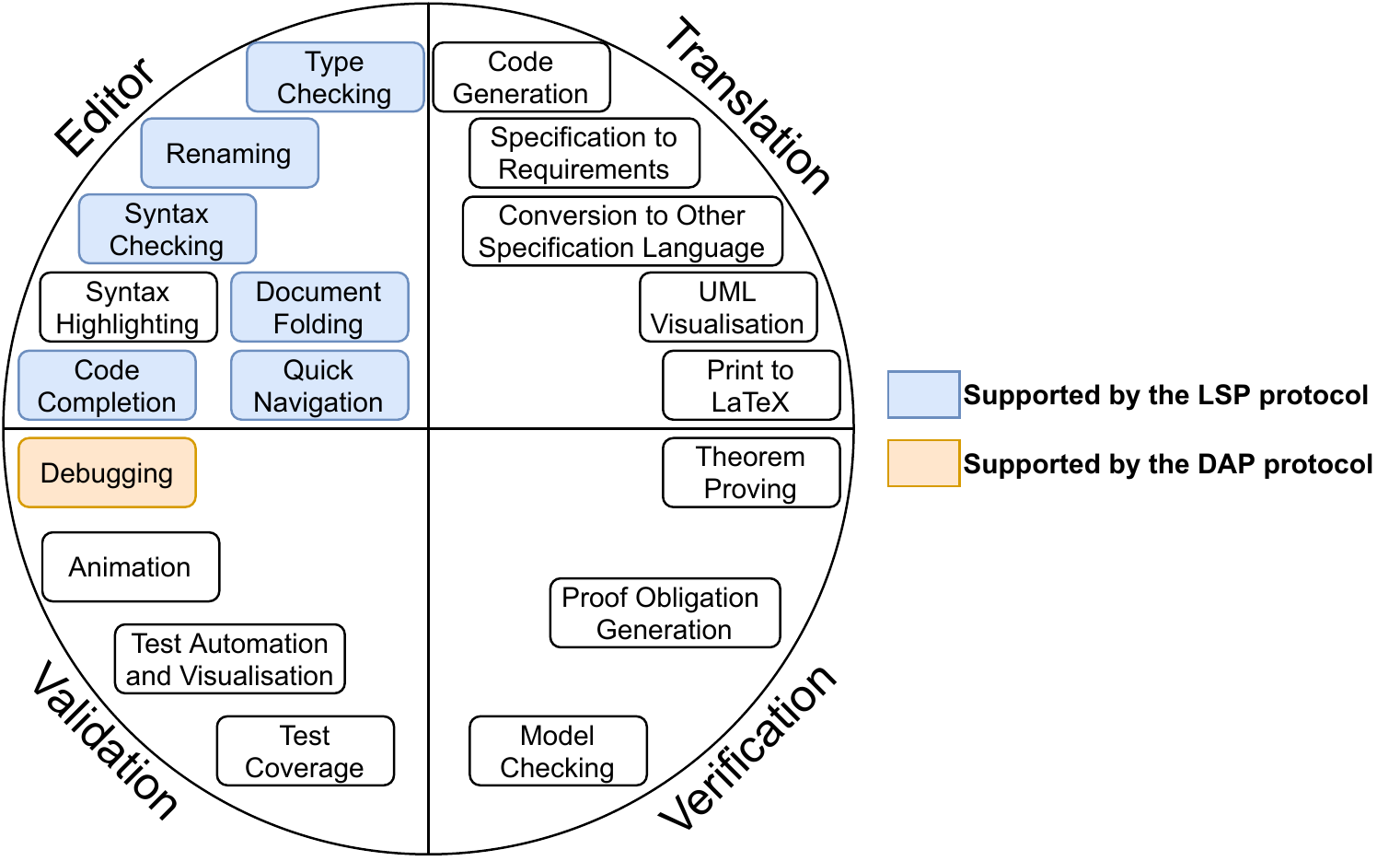}
    \caption{Specification language features covered by existing language-neutral protocols. The individual colours denotes which protocol is needed between the client and language server to support the given feature. Thus, for the LSP protocol a feature is coloured blue, while it is orange for the DAP protocol.}
    \label{fig:ProtocolFeatureCoverage}
\end{figure}

The protocols, LSP and DAP, have been developed for programming languages.
However, they can also be used with specification languages to support some of the features normally available to programming languages.
An overview of common features available for specification languages is illustrated in \Cref{fig:ProtocolFeatureCoverage}.
%The figure is based on 

To support editor features not including syntax highlighting a direct implementation of the \gls{lsp} protocol can be used to decouple the \gls{ide} from the language core that supports these features.
Similarly, the \gls{dap} protocol can be used for the debugging feature as it supports common debug functionality such as different types of breakpoints and variable values.
However, as clearly illustrated in \Cref{fig:ProtocolFeatureCoverage} many of the features that are not commonly found for programming languages cannot be supported using the existing protocols.
Thus, there is a need for further protocol development to support all the specification language features.

\section{Specification Language Server Protocol}
\label{sec:slsp}
% HDM: I am missing how we achieve what we mention in the related work. Somehow a good explanation how our extension to LSP is not add-hoc, and the LSP client class extension works...
To enable support for specification language features, we propose to use the \gls{slsp}. 
The protocol builds on the design and specification of the LSP protocol by extending the LSP protocol with language independent method calls and data-types.
One of the reasons for extending the LSP protocol instead of starting from scratch is that the LSP protocol is the de-facto standard for programming languages \cite{Rodriguez&18}. 
Furthermore, it provides the following benefits:
\begin{itemize}
    \item LSP defines a base protocol using three message types: \textit{request}, \textit{response} and \textit{notification}.
    
    \item LSP includes a number of basic structures than can be used to define messages, such as progress notification and cancel requests.
    
    \item LSP defines an infrastructure for communication initialisation between client and server. This enables clients and servers to exchange capabilities, which in turn allows the protocol to be used even if the client or server does not support all the protocol features. 
    
    \item LSP defines a synchronisation scheme that allows the server to have up-to-date project files even if the user changes them on the client side.
    
    \item Finally, the LSP protocol provides support for the editor features found in \Cref{fig:ProtocolFeatureCoverage}. Thus, these features are directly available in the SLSP protocol.
\end{itemize}

However, there are a few downsides to extending the LSP protocol. 
That is, the initialisation and text synchronisation capabilities are non-optional to implement. 
The initialisation messages are very simple to implement. 
However, the synchronisation may be slightly over-complicated for some language features,
\eg translation that just needs to know the state of the files when the user activates the features, thus it is not benefiting from the constant synchronisation of the files.

The protocol is developed with a focus on clarifying whether specification language features are able to be supported in a language neutral manner with a client and server communicating using a protocol similar to the \gls{lsp} protocol.
To this effect the initial set of specification language features supported by the protocol is primarily based on the feature support found in the Overture IDE \cite{Larsen01,Larsen&10a} for the dialects of \gls{vdm} for which the authors are proficient.
For this reason the features and messages may not be expressive enough for some specification languages. 
However, the protocol is easily extendable with optional entries or additional messages to support missing features.
How the protocol should be changed to fully support other languages is left as future work that should be decided on the inputs and requirements from other language specialists.

In the following subsections we describe how the features \gls{pog}, \gls{ct}, Translation and \gls{tp} are supported in the \gls{slsp} protocol to support VDM.
However, the protocol is based on language-neutral data types, which makes it applicable to any text based language.
An overview of the messages in the SLSP protocol is available in \Cref{tab:SLSPMessages}.
A detailed description of each of the messages in the SLSP protocol and the design decisions is available in \cite{Rask&21}.
Furthermore, the support for the features \gls{pog}, \gls{ct} and Translation have been tested in a pilot study described in \Cref{sec:pilotStudy}, however the support for \gls{tp} has not been tested.

\subsection{Proof Obligation Generation}
In order for the client to request \glspl{po} present in the specification, from the server, the message `generate' is defined in the protocol which enables the server to respond with a list of \glspl{po}.
A \gls{po} is defined by the type \texttt{ProofObligation} which contains an id, a name, \gls{po} type, location in the specification where the \gls{po} applies and an optional flag to indicate if the \gls{po} is proved.
Furthermore, a notification message is also defined which is used for synchronisation of the \glspl{po} with respect to the specification. 
This enables the client to request new \gls{pog} if the specification has changed.

\begin{figure}[htbp]
    \centering
    \includegraphics[width=0.8\textwidth]{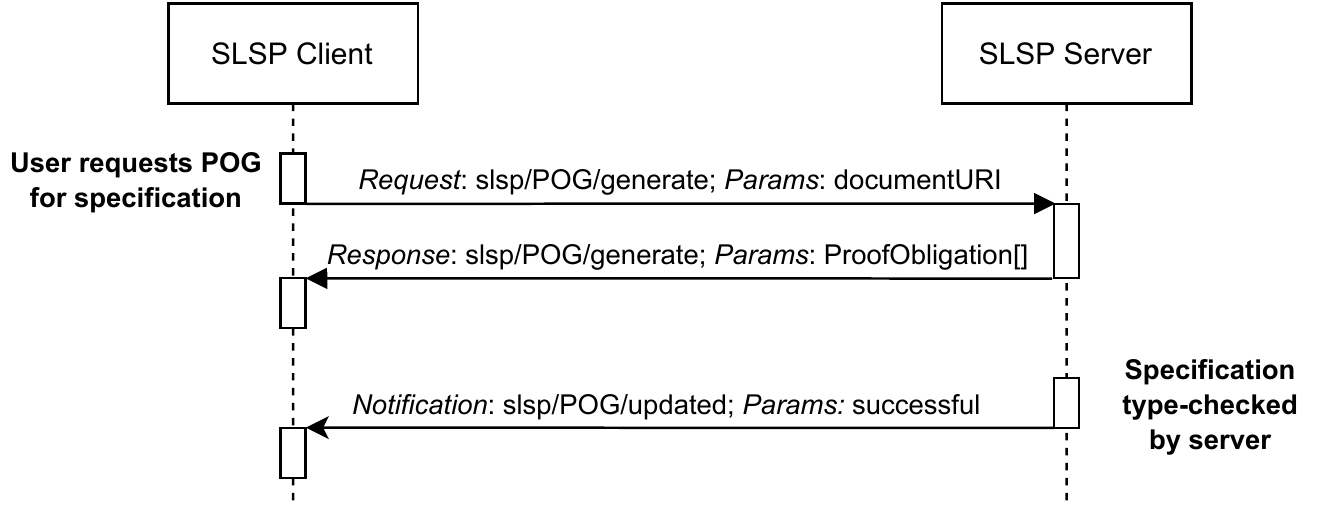}
    \caption{Example of how a client and a language server communicates during \acrshort{pog}.}
    \label{fig:pogSequenceDiagram}
\end{figure}

\Cref{fig:pogSequenceDiagram} illustrates an example of how a client and a language server communicate during a routine \gls{pog} session.
Similar sequence diagrams for the other features are available in \cite{Rask&21}.

\subsection{Combinatorial Testing}
For the client to be able to request traces present in the specification the message `traces' is defined in the protocol.
The server is able to respond to this message with a list of the type \texttt{CTSymbol} which contains the information necessary for building a tree structure of the traces present in the specification. 
This includes the name of the trace group, e.g. `classA' and a collection of traces for the group each specified by a fully qualifying name, their location in the source code and an optional verdict, i.e. if all tests for the trace are passed or failed.

The message `generate' allows the client to request the server to generate tests for a given trace to get an overview of the number of tests for the trace before execution.

The message `execute' enables the client to initiate test execution. 
The message requests execution of a given number of tests in a trace by providing the fully qualifying name of the trace, possible filtering options and the range of tests to execute.
The server is allowed to either make a single response containing all test results when test execution has finished or to 
respond with batches of test results asynchronously by sending partial results back to the client during test execution.
The response type is a \texttt{CTTestCase} which contains an id, a test verdict and the test case sequence and its result.

\subsection{Translation}
The protocol supports translation of a specification to any format supported by the server.
However, the translation must be able to be carried out purely with information available on the server side as only the single message `translate' is defined.

The message enables the client to request the server to initiate a translation of the specification to a given format by providing a location on the drive where the server is to store the resulting translation.
The server is able to respond with the location of a specific document of the resulting translation that should be opened in the editor.

\subsection{Theorem Proving}
Theorem Proving (TP) features exists for many specification languages, such as PVS\footnote{See \url{https://github.com/nasa/vscode-pvs}.}~\cite{Masci&19}, the Isabelle\footnote{See \url{https://marketplace.visualstudio.com/items?itemName=makarius.Isabelle2020}.}~\cite{Paulson86_2} and Lean\footnote{See \url{https://github.com/leanprover/vscode-lean}.}~\cite{Moura&15}.
However, such functionality is not yet available for VDM.
Hence, the support for TP in the SLSP protocol is largely based on the functionality available for PVS in the VSCode-PVS extension~\cite{Masci&19}.

To be able to prove a lemma it is expected that an overview of lemmas in the corresponding specification is available.
To this effect the message `lemmas' has been defined in the protocol, allowing the client to query the server for lemmas in a given specification.
Furthermore, to represent a lemma the type \texttt{Lemma} has been defined. 
This type contains its name, a name of the theory that the lemma belongs to, its location, its kind and its status.

To start the proving of a lemma the message `beginProof' is defined which allows the client to request the server to initiate the theorem prover for the requested lemma.
The server response is the initial state of the proof represented by a \texttt{ProofState} type which contains an id a status, potential sub-goals and rules.

Usage of \gls{atp} is handled in the protocol with the request message `prove'.
The message is send by the client to the server to request the theorem prover to automatically prove a lemma.
If a proof has been started the prover should attempt to find a solution for the current lemma at the current proof step.
Alternatively, the entry \texttt{name} is defined in the message which can be used to specify the name of a lemma that \gls{atp} should be applied to. 
If the \gls{atp} process needs to be cancelled before completion the \gls{lsp} cancel message is used, the same applies for cancelling other commands.
Furthermore, the command for initiating \gls{atp} often varies between theorem provers.
By defining a message for \gls{atp} in the protocol specification a single standardised interface, effectively moving the responsibility of calling the specific \gls{atp} command(s) to the server.
The response to the prove request is of the type \texttt{TPProveResponse} which contains a status for the proof, processing time, a list of suggested commands if any and a description of potential counter examples, proof steps, etc.

\begin{table}[htbp]
\centering

\begin{tabular}{p{0.155\textwidth}p{0.265\textwidth}p{0.25\textwidth}p{0.23\textwidth}}
\hline\hline
\multicolumn{1}{c}{\textbf{Method}} & \multicolumn{1}{c}{\textbf{Description}}                                                           & \multicolumn{1}{c}{\textbf{Parameters}}                                                                   & \multicolumn{1}{c}{\textbf{Response}}                           \\ \hline
POG/generate                          & \begin{tabular}[c]{@{}l@{}}Generate proof obligations \\ for a file or folder\end{tabular}          & uri: DocumentUri                                                                                           & ProofObligation{[}{]}                                           \\ \hline
POG/updated                           & \begin{tabular}[c]{@{}l@{}}Notify that the specification \\ has been updated\end{tabular}           & successful: boolean                                                                                        & null                                                             \\ \hline
CT/traces                             & \begin{tabular}[c]{@{}l@{}}Request an outline of the \\ traces in the specification\end{tabular}    & uri?: DocumentUri                                                                                          & CTSymbol{[}{]}                                                   \\ \hline
CT/generate                           & Generate tests for a trace                                                                          & name: string                                                                                               & \begin{tabular}[c]{@{}l@{}}numberOfTests: \\ number\end{tabular} \\ \hline
CT/execute                            & \begin{tabular}[c]{@{}l@{}}Execute tests for a trace,\\ with optional filter and range\end{tabular} & \begin{tabular}[c]{@{}l@{}}name: string\\ filter?: CTFilterOption{[}{]}\\ range?: NumberRange\end{tabular} & CTTestCase{[}{]}                                                 \\ \hline
TR/translate                          & \begin{tabular}[c]{@{}l@{}}Translates the specification \\ into the specified language\end{tabular} & \begin{tabular}[c]{@{}l@{}}uri?: DocumentUri\\ languageId: string\\ saveUri: DocumentUri\end{tabular}      & uri: DocumentUri                                                 \\ \hline
TP/lemmas                             & \begin{tabular}[c]{@{}l@{}}Request list of lemmas in \\ the specification\end{tabular}              & projectUri?: DocumentUri                                                                                   & Lemma{[}{]}                                                      \\ \hline
TP/beginProof                         & \begin{tabular}[c]{@{}l@{}}Initiate a proof session for \\ a lemma\end{tabular}                     & name: string                                                                                               & ProofState                                                       \\ \hline
TP/prove                              & \begin{tabular}[c]{@{}l@{}}Automatically prove \\ lemma or request hint\end{tabular}                & name?: string                                                                                              & TPProveResponse                                                  \\ \hline
TP/getCommands                        & \begin{tabular}[c]{@{}l@{}}Get list of commands that \\ can be applied to the proof\end{tabular}    & null                                                                                                       & TPCommand[]                                                        \\ \hline
TP/command                            & \begin{tabular}[c]{@{}l@{}}Apply command to the \\ current step of a proof\end{tabular}             & command: string                                                                                            & TPCommandResponse                                                \\ \hline
TP/undo                               & Undo a proof step                                                                                   & id?: number                                                                                                & ProofState                                                       \\ \hline\hline
\end{tabular}
\caption{List of new messages available in SLSP, for more information see \cite{Rask&21}. All methods are prefixed with \texttt{slsp/}, \eg \texttt{slsp/POG/generate}.
A `?' after the name of an entry indicates that the entry is optional.
}
\label{tab:SLSPMessages}
\end{table}

To facilitate \gls{itp} it should be possible to send commands to a prover at the server side.
This is possible with the message `command' in the protocol.
The server should respond with the type \texttt{TPCommandResponse} which contains a description of the result of the command as a human-readable string and a proof state of the type \texttt{ProofState}.
To get the list of available prover commands the message `getCommands' is defined in the protocol. 
The response is a collection of type \texttt{TPCommand} which contains a name and a description of the command.

To undo a proof step the message `undo' is defined with the response of type \texttt{ProofState} which enables the server to respond with the previous state of the proof.
For undoing a specific step that is not the latest one the message includes an \texttt{id} entry that specifies the id of the step that the theorem prover should undo.
To step through a proof the \texttt{command} request should be used.
Thus, stepping through a proof is simply a re-transmission of previous prover commands, one for each step.

As specifications and lemmas can change after they have been proved, it is relevant to be able to re-run a proof.
One example of enabling this functionality is to have the client store all commands that have been transmitted to complete a given proof, and keep them stored even if the specification changes.
Thus the client is able to re-transmit all the stored commands and check if the proof is still completed.
Alternatively, commands could be stored in the source text removing the need for the client to keep a state relating to commands while also alleviating the need to communicate the commands directly to the server as this would be done indirectly through the source text.

\FloatBarrier

\section{Pilot Study}
\label{sec:pilotStudy}
% Jonas
% \begin{itemize}
%     \item Shortly about VDM-VSCode
%     \item functionality comparison between Overture and VDM-VSCode
%     \item LoC, argue for similar or less effort when using protocol compared to language specific API
% \end{itemize}

To test the \gls{slsp} protocol we have carried out a pilot study\footnote{The source code for VDM VSCode is found at: \url{https://github.com/jonaskrask/vdm-vscode}.}, that implements support for the specification language \gls{vdm}, more precisely the dialects VDM-SL, VDM++ and VDM-RT.
It is implemented in \gls{vscode} as this already supports the protocols \gls{lsp} and \gls{dap}.
A detailed description of the support for VDM in \gls{vscode} has previously been published \cite{Rask&20}.
This paper will provide a short description of the extension, VDM VSCode, which is used as a basis for evaluating the use of the language-neutral protocols.
The evaluation focuses on the implementation efforts related to using the language-neutral protocols compared to a more language-specific implementation that directly access the language support functionality.

\subsection{VDM VSCode}
\begin{figure}[htb]
    \centering
    \includegraphics[width=\textwidth]{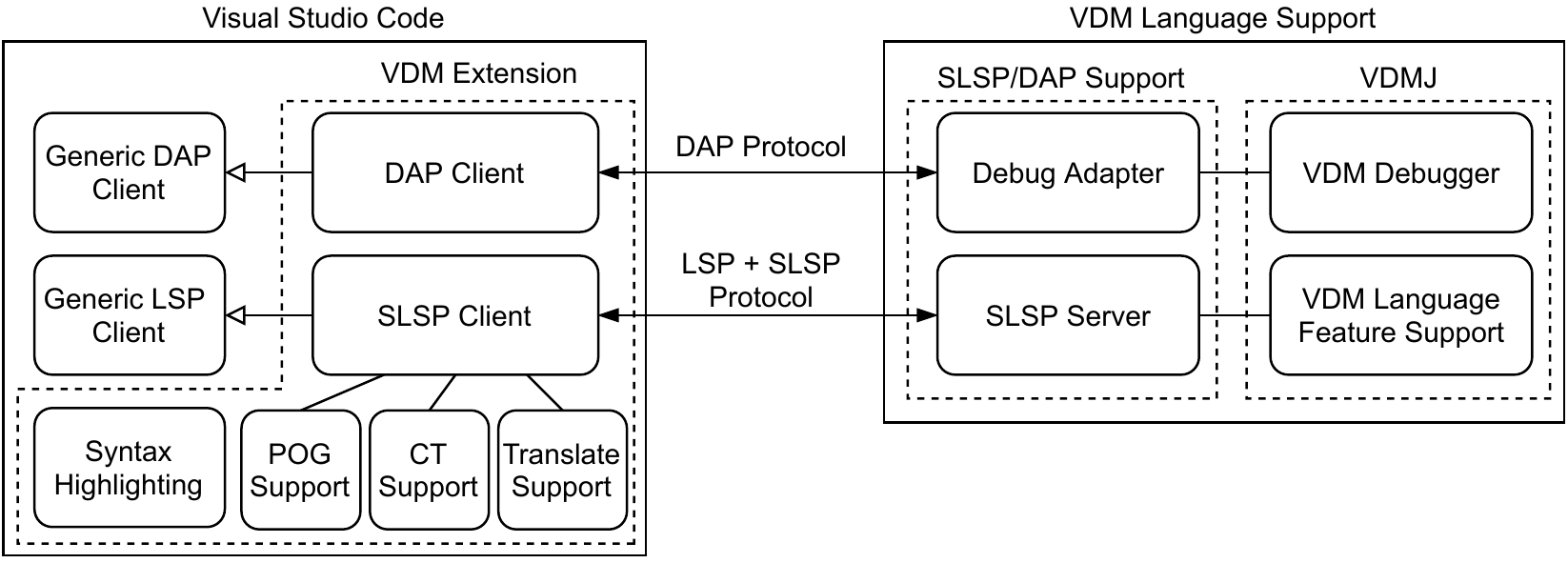}
    \caption{Overview of the architecture of the VDM language extension for VS Code.}
    \label{fig:ExtensionArchitecture}
\end{figure}

To support VDM in VS Code an extension is developed which implements a client-server architecture that uses the SLSP, LSP and DAP protocols for communication, as illustrated in \Cref{fig:ExtensionArchitecture}.

The language support is provided by the server that builds on the VDM language core: VDMJ~\cite{Battle09}.
The functionality of VDMJ is exposed to the protocols through a separate jar that enables the protocol communication and performs function calls to VDMJ.
To handle the DAP protocol a debug adapter is implemented that translates DAP messages into actions in the VDM debugger and vice versa.
To handle the SLSP and LSP protocol, a SLSP server is created that amongst other things handles the synchronisation of project files such that the client and server always agree on the state of a project. 
Furthermore, the server handles LSP and SLSP messages and translates these into language actions that are executed using the VDMJ functionality.

\begin{figure}[htb]
    \centering
        \includegraphics[width=1\textwidth]{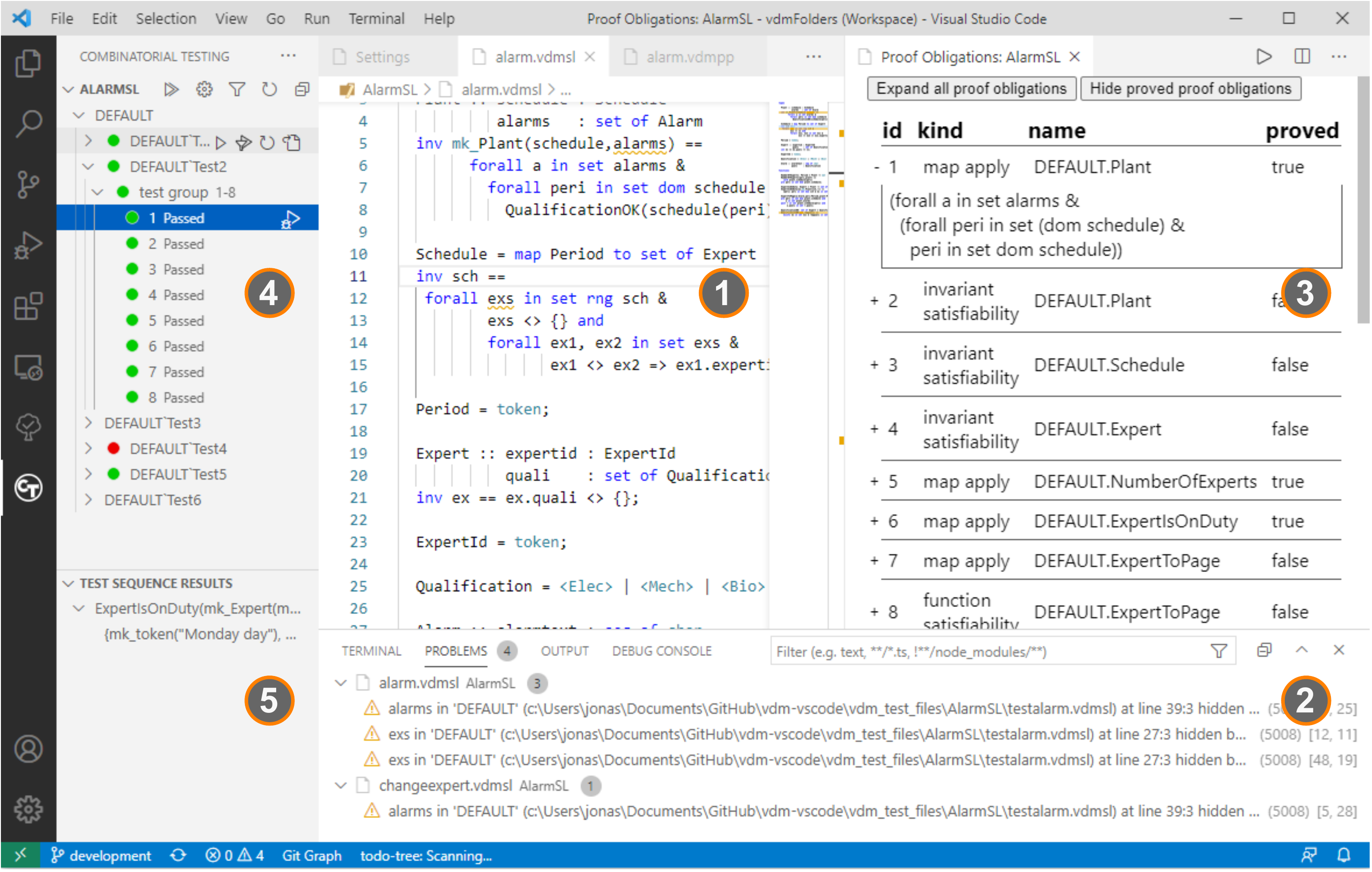}
    \caption{Screenshot of the VDM VSCode Extension: (1) Main Editor; (2) Problems View; (3) Proof Obligation View; (4) Combinatorial Testing View; (5) Test Result View}
    \label{fig:VDMVSCodeScreenshot}
\end{figure}

The client implementation utilise the VS Code API to provide support for the DAP and LSP protocols.
Both of these are supported by generic client modules that provide full support for the protocols with very little extra code.
The generic LSP client is easily expandable which allows the SLSP client to reuse the basic message infrastructure.
To support the features of SLSP new language-agnostic modules are created that provide message handling and GUI elements for each feature.
For the POG feature a table-like view is implemented to show proof obligations and meta-data.
For the CT feature a tree view is implemented for navigation between tests and traces.
Since the SLSP modules are language-agnostic anyone who adhere to the SLSP protocol will be able to reuse the modules to support any other language.
The new views from the VDM VSCode extension are illustrated in \Cref{fig:VDMVSCodeScreenshot}.

\subsection{Implementation Effort} 
A concern that might arise, when considering to use the SLSP protocol for the language support, is related to the implementation efforts required.
A Lines of Code (LoC) measurement is used to quantify the implementation effort for a given language feature which enables a comparison between VDM VSCode (v1.1.0) not including the server and the Overture IDE (v3.0.1) not including the language core, this is found in \Cref{tab:LocComparison}. 
To be able to properly compare the two IDEs an analysis of the functionality available for each feature is conducted, which found that VDM VSCode has the same or higher level of support for the implemented features.

\begin{table}[htb]
\centering
\begin{tabular}{|l|r|r|}
\hline
\textbf{Feature} & \textbf{VDM VSCode} & \textbf{Overture IDE} \\ \hline
Editor               & 383     & 10,163     \\ \hline
% Syntax Highlight     & 960     & 671        \\ \hline
Debug                & 60      & 22,865     \\ \hline
POG                  & 494     & 711        \\ \hline
CT                   & 1,014    & 4,727       \\ \hline
Translate to LaTeX   & 100     & 1,766       \\ \hline
Translate to Word    & 2       & N/A        \\ \hline
\textbf{Sum}         & \textbf{2,068} & \textbf{40,232} \\ \hline
% \textbf{Sum}         & \textbf{3028} & \textbf{40,903} \\ \hline
\end{tabular}
\caption{Comparison of LoC\protect\footnotemark  (excluding comments and blank spaces) of language feature implementations in VDM VS Code not including the server and the Overture IDE not including the language core. N/A indicates that the feature is not present in the IDE.}
\label{tab:LocComparison}
\end{table}

From the comparison it is clear, by the sum of LoC, that VDM VSCode requires significantly less LoC on the IDE side than the Overture IDE, with the most significant differences for the editor and debug features.
In VDM VSCode these features are supported by the LSP and DAP protocols, respectively.
This is made possible due to the generic support for the LSP and DAP protocols available in the VS Code API.
Thus, there is a significant reduction in implementation effort when providing support for the features that are generically supported by the IDE, made possible by the language-agnostic nature of the protocols.
But what about the new SLSP features that are not supplied with the LSP support?

\footnotetext{The LoC is counted using the `VS Code Counter' extension found at: \url{https://marketplace.visualstudio.com/items?itemName=uctakeoff.vscode-counter}}

The features that are supported by the SLSP protocol are the POG, CT, and translation features.
For POG the implementation effort for VDM VSCode is slightly less than that of the Overture IDE. 
In VDM VSCode about half of the LoC are attributed to the protocol handling and the other half to the GUI elements.
For CT the LoC are approximately twice of the POG ones, this is expected as the CT feature includes a lot more functionality than the POG feature.
Compared to the Overture IDE the effort is significantly reduced.
The same is true for the translation features.
Thus, it seems that it does not require additional implementation effort to implement the language features in a language-agnostic manner, using the protocols, compared to accessing a more language-specific API, such as that provided by the Overture language core.
Furthermore, it should be noted that the POG, CT, and translation modules developed for VDM VSCode are language-agnostic as a result of using the SLSP protocol.
That means that any language server that implements the SLSP protocol, can use the modules with very little effort, as evident from the efforts related to the editor and debug features.

Since the language support is divided into a client and a server, implementation effort is also required on the server side to expose the language features through the protocol interface.
If you are developing a new language core, the protocol interface can be incorporated from the beginning.
However, if a language core is already available this is extra work that must be carried out to use the protocols to provide the language support.
Thus, for a complete picture of the effort to implement support for a specification language using the SLSP protocol, the LoC count for the server-side implementation should also be considered.
However, it is important to note that this implementation effort only has to be made once for the server to be available across all IDEs that implement support for the protocols.

\subsubsection{Effort Related to the Server}
One important aspect of implementing the pilot study was to discover how closely the programming language concepts in the LSP and DAP protocols match the concepts in the VDM specification language. 
The concepts in the protocols are deliberately abstract, such as a location within a file or a symbol name, rather than being language specific. The problem is to map these abstract concepts into concrete language features in the chosen VDM dialect.

The most significant difference for VDMJ operating via LSP is that the file system does not necessarily represent the latest version of a source file. Rather, the client can send `didChange' events at any time to indicate that a file is actively being edited. These changes must be syntactically checked on the fly, and the user is under no obligation to save the files to disk before seeing any errors. This clashes with the design of the VDMJ type checker, which needs to see the entire specification. So the LSP server design caches user changes in memory. 

Several of LSP's language features map well to VDM. However, one annoyance with outline information is that symbols have to be one of a fixed set of kinds, like `class', `method', `interface' and so on. The set is reasonably rich, but there is no natural mapping for some of VDM's definition categories - for example what is a combinatorial trace, or a permission predicate guarding an operation?

\begin{table}[htb]
\centering
\begin{tabular}{|l|l|r|}
\hline
\textbf{Package} & \textbf{Description} & \textbf{LoC} \\ \hline
json &		JSON message support 	& 787  \\ \hline
rpc	 &		JSON RPC support		& 284  \\ \hline
lsp	 &		LSP/SLSP protocol support	& 2,033 \\ \hline
dap  &		DAP protocol support	& 1,265 \\ \hline
vdmj &		VDMJ interface support	& 3,374 \\ \hline
workspace &	Workspace coordination	& 3,342 \\ \hline
\textbf{Sum}         &  & \textbf{10,985} \\ \hline
\end{tabular}
\caption{Java LoC for the server (excluding comments and blank spaces) not including the VDMJ language core. }
\label{tab:LocServer}
\end{table}

\Cref{tab:LocServer} contains an overview of the implementation efforts of extending VDMJ to handle the SLSP protocol indicated as lines of Java code.
The support consists of six packages.
The packages `json' and `rpc' handle the basic message support, with no knowledge of the SLSP, LSP and DAP protocols, hence you could use those packages to write an arbitrary JSON RPC server 
The `lsp' and `dap' packages contain handlers, for the SLSP, LSP and DAP protocol messages, that delegate to the workspace to get the job done.
Lastly, the workspace package understands `how' to do things, handles all the VDM sources, and uses the `vdmj' package to talk to VDMJ.

It should be noted that there exists several SDKs that implement the LSP and DAP protocols\footnote{See \url{https://microsoft.github.io/language-server-protocol/implementors/sdks/}.}, \eg LSP4J\footnote{See \url{https://github.com/eclipse/lsp4j}.} that makes it easy to implement a language client and server in Java.
Using such an SDK would reduce the efforts as they supply the functionality of the `json', `rpc' and `dap'.
Also, it will implement most of the `lsp' package, except the SLSP parts.
However, since no such SDK was used for the server it is unknown whether it could be extended for SLSP easily.

\subsubsection{Combining the Efforts}
Combining the LoC for the client and server the VDM VSCode extension requires a total of $2068 + 10,985 = 13,053$ LoC.
This is still significantly less than the Overture IDE that requires a total of 40,903 LoC.
Based on these results, it is expected that the implementation efforts to support future features of the SLSP protocol are less than that of their comparable implementation in the Overture IDE.
Similarly, it is expected that feature support for other languages will not require an additional implementation effort by using the protocols compared to more direct support.
Thus, the tool and language developers will be able to benefit from the long term advantages of having separate language-agnostic clients and language-specific servers, without an increased initial implementation effort.

\section{Related Work}
\label{sec:relatedWork}
% Frederik will write
% Hugo will read through 2nd round

The interest of having a protocol facilitated language support for other formal languages besides \gls{vdm} is evident in tool implementations for related formal specification languages such as \gls{pvs} \cite{Masci&19}, Dafny \cite{Hess&19}, Coq\footnote{See \url{https://github.com/siegebell/vscoq}.} \cite{Dowek&93}, Isabelle\footnote{See \url{https://marketplace.visualstudio.com/items?itemName=makarius.Isabelle2020}.} \cite{Paulson86_2} and Alloy\footnote{See \url{https://github.com/s-arash/org.alloytools.alloy/tree/ls}.}. %The citations for Coq and Isabelle are not related to VS Code or \gls{lsp} so should they be removed?
A common pattern in all the implementations is the leverage of the \gls{lsp} protocol for editor-related features. 
Some of them also include language features beyond the scope of the \gls{lsp} protocol, for which they implement their own language-specific protocols.
This is done by extending the \gls{lsp} protocol with language specific messages resulting in a protocol that is no longer language-agnostic.

This is also present in the paper by Masci and Muñoz~\cite{Masci&19}. From the architecture it is found that the extension integrates the editing and proof management functionalities of \gls{pvs} in VS Code, enabling functionalities that developers expect to find in modern verification tools.
To enable support for features that are not supported by the \gls{lsp} protocol, such as interactive analysis of \gls{pvs} specifications, the protocol is extended resulting in an extended protocol that is not language-agnostic, because the extensions are bound to specific language features of \gls{pvs}. 
This allows their language server to be partially reused for \glspl{ide} that support the \gls{lsp} protocol.
However, support for the \gls{pvs}-specific protocol must be separately implemented. 

On the other hand, the \gls{slsp} protocol presented in this paper is developed with the purpose of supporting features that are common for multiple specification languages. For instance, \gls{slsp} supports proof obligation generation. Although its implementation was tailored to support the VDM feature needs, thus its naming; the protocol intends to cover the needs of extensions for languages generating proof goals. One such example  are the type-checking conditions generating feature available in the PVS VS Code extension.  
As such, the development of the \gls{slsp} protocol aims to increase the motivation for tool developers to add native support for the protocol, thus reducing the combined effort of implementing language support for specification languages in multiple \glspl{ide}.
That is why we believe that it is worth exploring the feasibility of using the \gls{slsp} protocol in tool implementations for specification languages that already leverages the \gls{lsp} protocol. 

\section{Concluding Remarks and Future Work}
\label{sec:conclusion}
\begin{figure}[htb]
    \centering
    \includegraphics[width=.85\textwidth]{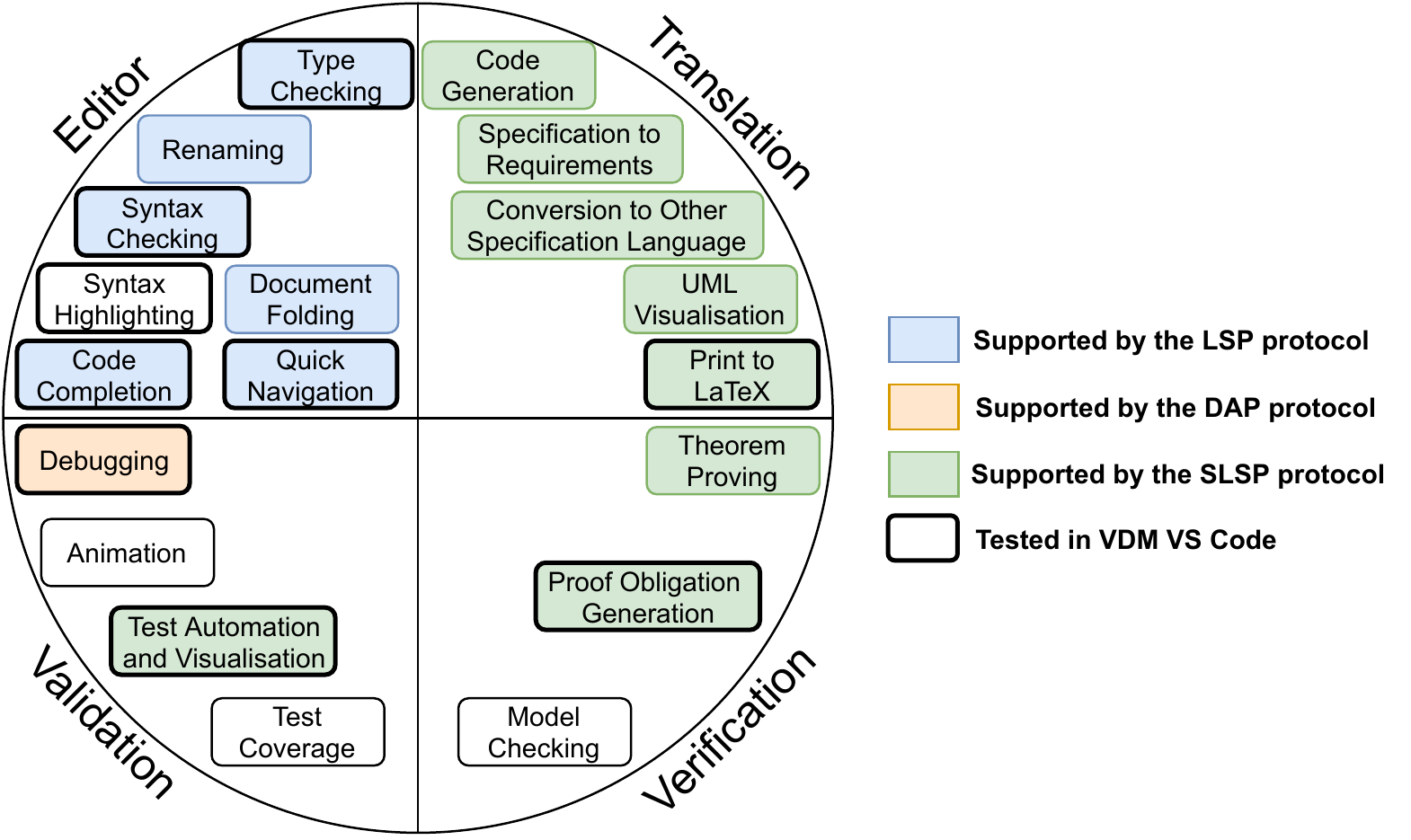}
    \caption{Specification language features supported by the LSP, DAP and SLSP protocols. A boldface edge indicates that a feature has been successfully implemented in the pilot study and is available in the VDM VS Code extension. The individual colours denotes which protocol is needed between the client and language server to support the given feature. Thus, for the LSP protocol a feature is coloured blue, while it is orange for the DAP protocol and green for the SLSP protocol.}
    \label{fig:SLSPFeatureCoverage}
\end{figure}

Our research on how to best extend the LSP and DAP protocols with the VDM suite of specification languages features lead to a new protocol proposal, the SLSP protocol. We implemented the first client and server for it, and it includes the non-programming language features of the VDM specification languages, and together with a LSP and DAP  protocol implementation, it extends VS Code to become an IDE for our specification languages and to be our future default implementation supporting them. 
An overview of our implementation coverage of specification language features is depicted in Figure~\ref{fig:SLSPFeatureCoverage}, where we assert which features are covered by each protocol, and which features that have been tested in the VDM VS Code extension.  

Our results show that the language-neutral support for specification languages is feasible, given the ability to extend LSP within our proof of concept. Nevertheless, our work is in its initial stages and  more work is needed to fully cover the needs of the VDM language. For example, we did not implement theorem proving in the proof of concept although it is supported by the SLSP protocol as reported in Figure~\ref{fig:SLSPFeatureCoverage}. 
By using LSP, DAP and SLSP, we reduced implementation effort, which will be  valuable in the long run, especially when extending other IDEs, as observed in the reduced number of LoC. In addition, the SLSP protocol was designed to be an extension to LSP to support specification language features in a decoupled manner, thus transforming the $M \times N$ problem into $M + N$ as LSP does.  
We expect our work motivates the other formal specification language community members to join efforts to develop a standardised  version of the SLSP protocol that adequately supports the language features found across various formal specification languages, thus extending this result to all of them.

%\subsection{Future Work}
% \begin{itemize}
%     \item Use protocol for other languages
%     \item "Super client" using multiple specialised servers to provide language support. Facilitated using the SLSP protocol.
%     \item Parallel execution of some tasks (e.g. CT)
%     \item Distributed support for specification languages
% \end{itemize}

The intention is that when SLSP is completed with the remaining Overture/VDM features, we expect this method to be the one that the VDM community is going forward with. However, we expect that the first step is that SLSP is used for the proof support parts of another specification language. \eg maturing the protocol to be able to support either Isabelle or Coq; some of the hardest interactive theorem provers to provide an interface.

Moving forward, it is our hope that the SLSP protocol becomes the standard for IDE support of specification language features, such as the LSP protocol is for editor features.
To get there, the SLSP protocol has to be tested for other specification languages than VDM through further pilot studies.
This will mature the protocol and identify any shortages in it, \eg for functionality not found in VDM that is available in other languages.

Another interesting path for future work is to utilise the introduction of a client-server architecture for the language support to enable distributed language support.
Since the base-protocol is very lightweight a client could be connected to multiple servers, each specialised for a specific feature, \eg one which is fast at small editorial actions that requires a quick response and another for long-running tasks such as theorem proving.
This could also be used for parallel execution of some tasks, \eg combinatorial testing would be able to benefit from this as each test is independent.

\section*{Acknowledgement}

We acknowledge the Poul Due Jensen Foundation for funding the project Digital Twins for Cyber-Physical Systems (DiT4CPS), and we would like to thank the reviewers for their thoughtful comments and suggestions on the SLSP protocol.
% \nocite{*}
\bibliographystyle{eptcs}
% \bibliography{generic}
\bibliography{refs.bib, dan.bib}

\end{document}